\documentclass[12pt]{article}

\newcommand{\Hc}{\mbox{H.c.}}
\newcommand{\ol}{\overline}
\newcommand{\norm}[1]{\left|#1\right|}

\newcommand{\W}{{\cal W}}

\newcommand{\M}{M_{\rm R}}

\newcommand{\eq}[1]{(\ref{#1})}
\newcommand{\Eq}[1]{Eq.~(\ref{#1})}
\newcommand{\eqs}[2]{(\ref{#1})--(\ref{#2})}

\newcommand{\Nc}{N_{{\rm c}}}
\newcommand{\Nf}{N_{{\rm f}}}
\newcommand{\Nm}{N_{{\rm m}}}
\newcommand{\fun}[1]{\!\left(#1\right)}
\newcommand{\abs}[1]{\left\vert#1\right\vert}
\newcommand{\massF}{B_{\mathrm{R}}}
\newcommand{\massD}{Y_{\mathrm{R}}}

\renewcommand{\thanks}{\footnote}
\newcommand\tocite[1]{%
${\hbox{--}}$\cite{#1}}

\makeatletter
\def\@JLone<#1,#2>{#1}
\def\@JLtwo<#1,#2,#3>{#2}
\def\@JLyear<#1,#2,#3,#4>{#3}
\def\@JLpage<#1,#2,#3,#4>{#4}
\newcommand\JL[1]{\@JLone<#1>\ {\bfseries \@JLtwo<#1>} (\@JLyear<#1>), \@JLpage<#1>}
\def\@Jpage<#1,#2,#3>{#3}
\newcommand\andvol[1]{{\bfseries \@JLone<#1>} (\@JLtwo<#1>), \@Jpage<#1>}

\newcommand\PTPS[1]{Prog. Theor. Phys. Suppl.\ {No. \@JLone<#1> (\@JLtwo<#1>), \@Jpage<#1>}}
\makeatother

\newcommand\PRD[1]{Phys.\ Rev.\ D\ \andvol{#1}}

\newcommand\PLB[1]{Phys.\ Lett.\ B\ \andvol{#1}}

\newcommand\NPB[1]{Nucl.\ Phys.\ B\ \andvol{#1}}

\newcommand\JHEP[1]{J. High Energy Phys.\ \andvol{#1}}

\begin{document}

\begin{titlepage}
\vspace*{1cm}
\begin{center}
{\huge\bf
 Soft Supersymmetry Breaking 
\\[0.2cm]
at Heavy Chiral Threshold
}

\vspace*{1.5cm}
{\large
Hiroyuki~\textsc{Matsuura}\rlap,\,\footnote{
	E-mail: matsuura@muse.sc.niigata-u.ac.jp}
Hiroaki~\textsc{Nakano}\,\footnote{
	E-mail: nakano@muse.sc.niigata-u.ac.jp}
and
Koichi~\textsc{Yoshioka}\,\footnote{
	E-mail: yoshioka@higgs.phys.kyushu-u.ac.jp}%
}
\vspace*{5mm}

$^{1,2}${\it
Department of Physics, Niigata University, Niigata 950-2181, Japan%
}

$^3${\it
Department of Physics, Kyushu University, Fukuoka 812-8581, Japan%
}

\vspace*{1cm}
\vspace*{0.5cm}

\begin{abstract}
We discuss the structure of threshold corrections 
to soft supersymmetry-breaking parameters at the mass threshold 
of heavy chiral superfields. 
Nontrivial dependence on soft parameters of heavy matter fields
originates from the `physical' definition of the threshold scale,
at which the general form of soft supersymmetry breaking is derived
in the superfield coupling formalism.
\end{abstract}


\end{center}

\end{titlepage}

\setcounter{footnote}{0}

In theories with softly-broken ${\cal N}=1$ supersymmetry, 
the renormalization of soft supersymmetry breaking parameters 
can be described by extending 
the corresponding supersymmetry preserving coupling 
into superspace \cite{spurion,HisanoShifman}.
Such extension valid to all orders in perturbation theory 
is summarized by \cite{Deltaterm}\tocite{continuation}
\begin{eqnarray}
\ln \widehat\alpha\fun{\theta,\bar\theta}
   &=& \ln\alpha 
	+ \left(\theta^2 m_g + \Hc\right)
	+ \theta^2\bar\theta^2 X_g \ ,
	\qquad X_g = \norm{m_g}^2 + \Xi_g \ ,
\label{extended:gauge}
\\
\ln \widehat Z_i^{-1}\fun{\theta, \bar\theta}
  &=& \ln Z_i^{-1} 
	- \left(\theta^2 A_i + \Hc\right)
	+ \theta^2\bar\theta^2 m_i^2 \ ,
\label{extended:Z}
\end{eqnarray}
where $Z_i$ and $\alpha=g^2/8\pi^2$ are wavefunction renormalization 
with flavor index $i$ and gauge coupling constant respectively;
we also denote the corresponding gaugino mass $m_g$, $A$-type
parameter $A_i$ (which is not $A$-parameter itself)
and scalar mass $m_i^2$.
The $\theta^2\bar\theta^2$-component $\Xi_g$ of $1/\widehat\alpha$
plays a nontrivial role 
beyond one loop \cite{continuation,Deltaterm}.
The superpotential coupling ${\cal W} = y\phi_i\phi_j\phi_k$ 
also have a superspace extension
\begin{eqnarray}
\ln \widehat{y}\fun{\theta, \bar\theta}
	 = \ln y - \theta^2 {A_y}
	 + \frac{1}{2}\theta^2\bar\theta^2 {X_y} \ ,
\label{extended:Y}
\end{eqnarray}
where $A_y$ is the corresponding $A$-parameter
and $X_y = m_i^2 + m_j^2 + m_k^2$.

The superspace extension \eqs{extended:gauge}{extended:Y}
correctly incorporates the renormalization of divergences
in a softly-broken theory,
and thus one can extract the exact form of soft beta functions.
The purpose of this letter is to use this superspace extension
to extract 
the structure of threshold effects 
on soft supersymmetry breaking terms.
To this end, it is important to match the couplings 
at the `physical' mass threshold of heavy (chiral) fields.
We then argue that 
the mass threshold should be extended
into superspace as a real superfield.
By taking these points into account,
we derive a compact expression for the threshold effects
on soft terms in terms of supersymmetric anomalous dimensions.

In was shown \cite{GiudiceRattazzi,continuation} 
in the context of gauge-mediated supersymmetry breaking (GMSB) that, 
when the heavy mass threshold is approximately supersymmetric,
the 
soft terms can be calculated
by extracting the dependence of \textit{divergences}\/
in supersymmetric couplings on the threshold mass scale;
The finite threshold effects can be extracted
to the leading order in supersymmetry breaking.
To find a formula that is applicable 
to other types of supersymmetry breaking,
we generalize the method by combining it 
with the superspace extension of renormalized couplings.

To explain our idea, 
let us consider an $\mathrm{SU}(\Nc)$ supersymmetric QCD with 
$\Nf$ flavors of light matter fields $\phi_i$ and $\ol{\phi}_i$
($i=1,\cdots,\Nf$).
We assume that 
each matter field receives the wavefunction renormalization $Z_i$
without kinetic mixing.
In addition, the theory contains $\Nm$ copies of heavy matter fields
$\Phi_m$ and $\ol{\Phi}_m$ ($m=1,\cdots,\Nm$)
with the mass term ${\cal W}_{\rm mass} = M_{\rm H}\Phi\ol\Phi$.
In the following,
primed quantities represent those in the high-energy theory.
We also denote by $\Delta$ the difference at the mass threshold 
between an unprimed and primed quantities, e.g., 
$b=3\Nc-\Nf$, $b'=3\Nc-\left(\Nf+\Nm\right)$ and
$\Delta{}b\equiv{}b-b'\left(=\Nm\right)$
for the one-loop beta function coefficient.

An important point is that
the holomorphic mass parameter $M_{\rm H}$ is \textit{not}\/ really
the `physical' mass of the heavy particles;
the latter should be determined by properly taking into account 
the wavefunction $Z_\Phi'$ and $Z_{\ol\Phi}'$ 
of heavy matter fields $\Phi_m$ and $\ol{\Phi}_m$.
In fact, the proper mass threshold $\M$ at which
the renormalization-group evolutions should be matched is defined by 
a self-consistent relation \cite{matching,continuation}
\begin{eqnarray}
\M^2
  = \abs{M_{\rm H}}^2
	Z_\Phi^{\prime-1}\fun{\mu=\M}
	Z_{\ol\Phi}^{\prime-1}\fun{\mu=\M}
    \ .
\label{matching}
\end{eqnarray}
Note that the right-hand side is nothing but 
the running mass parameter $\M\fun{\mu}$ evaluated at $\mu=\M$.

We are interested in supersymmetry breaking effects
at the heavy mass threshold.
As in the GMSB case, the holomorphic mass parameter $M_{\rm H}$ 
can be extended as a chiral superfield according to
\begin{eqnarray}
M_{\rm H} 
	&\longrightarrow& 
	\widehat M_{\rm H} = M_{\rm H} - \theta^2 F \ .
\label{holo.th.}
\end{eqnarray}
In addition,
the wavefunction factors also bring
the supersymmetry breaking effects;
\begin{eqnarray}
\ln Z_\Phi'^{-1} &\longrightarrow& \ln\widehat Z_\Phi'^{-1} 
	= \ln Z_\Phi'^{-1} - [\theta^2 A_\Phi' + \Hc] 
	+ \theta^2\bar\theta^2 m_\Phi'^2 \ ,
\end{eqnarray}
and similar for $\ln Z_{\ol\Phi}'^{-1}$.
Correspondingly we introduce 
\textit{a real superfield extension} of the heavy mass threshold by
\begin{eqnarray}
\ln\widehat M_{\rm R}^2
	 = \ln\M^2 - [\theta^2 B_{\rm R} + \Hc] 
	 + \theta^2\bar\theta^2 Y_{\rm R} \ ,
\label{threshold}
\end{eqnarray}
where $B_{\rm R}$ and $Y_{\rm R}$ can be determined
by extending the self-consistent definition (\ref{matching})
into superspace.
Explicitly we find 
\begin{eqnarray}
B_{\rm R}
  &=& 	\frac{1}{1-\gamma'_M\fun{\M}}
	\left[\frac{F}{M_{\rm H}} + A_M'\fun{\M}\right] \ ,
\\
Y_{\rm R}
 &=& 
	\frac{1}{1-\gamma'_M\fun{\M}}
	\left[
	X_M'\fun{\M}
	+\abs{\massF}^2\dot{\gamma}_M'\fun{\M}
	+\left\{\massF^\dagger\dot{A}_M'\fun{\M}
	+\Hc
	\right\}
	\right]
	\ .
\end{eqnarray}
Here
$\gamma_M'\equiv\gamma_{\bar\Phi}'+\gamma_{\ol\Phi}'$ 
with $\gamma_{\Phi}'=d\ln Z_\Phi'^{-1}/d\ln\mu^2$
and $\gamma_{\ol\Phi}'\equiv d\ln Z_{\ol\Phi}'^{-1}/d\ln\mu^2$,
and the dot stands for $d/d\ln\mu^2$.
The $\theta^2$-component $B_{\rm R}$ 
is essentially the renormalized $B$-parameter, 
which is the sum of the bare $F/M_{\rm H}$ 
and the finite counterterms $A'_M\equiv{}A_\Phi'+A_{\ol\Phi}'$ 
corresponding to heavy chiral superfields.
The $\theta^2\bar\theta^2$-component $Y_{\rm R}$ 
contains $X'_M\equiv{}m'^2_\Phi+m'^2_{\ol\Phi}$, 
which plays an important role in calculation of soft scalar masses.

Once the superfield extension 
of the threshold mass, \Eq{threshold}, is determined,
it is straightforward to calculate 
the threshold effect on a 
supersymmetry breaking parameter 
from the matching condition 
for the corresponding supersymmetry preserving coupling.
Let us start with the matching condition for the gauge coupling constant
\begin{eqnarray}
\alpha(\M)
  &=& \alpha'(\M) \ .
\label{matching:gauge}
\end{eqnarray}
We substitute $\mu=\widehat{M}_{\mathrm{R}}$ 
into the superspace extensions
of $\alpha\fun{\mu}$ and $\alpha'\fun{\mu}$,
\begin{eqnarray}
\ln \widehat\alpha\fun{\mu}
   &=& 	\ln\alpha\fun{\mu} 
	+ \left[\theta^2 m_g\fun{\mu} + \Hc\right] 
	+ \theta^2\bar\theta^2 X_g\fun{\mu} \ .
\label{extended:gauge:again}
\end{eqnarray}
Then by requiring the superspace extended matching condition,
$\widehat\alpha(\mu=\widehat{M}_{\mathrm{R}})
  = \widehat\alpha'(\mu=\widehat{M}_{\mathrm{R}})$,
and taking the $\theta^2$-component of this equation,
we find the matching relation for the gaugino mass
\begin{eqnarray}
\Delta m_g
 \equiv m_g\fun{\M}-m_g'\fun{\M}
  = 	B_{\rm R}\Delta\gamma_\alpha \ .
\label{gaugino:tree}
\end{eqnarray}
Here $\gamma_\alpha\equiv{}d\ln\alpha/d\ln\mu^2$
is the gauge beta function $\beta$ divided by $2\alpha$.

Similarly, if we start from the matching condition
for the wavefunctions,
\begin{eqnarray}
\ln Z_i^{-1}(\M)
  &=& \ln Z_i'^{-1}(\M) \ ,
\label{matching:Z:tree}
\end{eqnarray}
we obtain for
$\Delta{A_i}\equiv{}A_i-A^{\prime}_i$ and
$\Delta{m^2_i}\equiv m^2_i-m^{\prime2}_i$ at $\mu=\M$,
\begin{eqnarray}
\Delta{A_i}\fun{\M}
 &=&{}-\massF\Delta\gamma_i \ ,
\label{Aterm:tree}
\\
\Delta{m^2_i}\fun{\M}
 &=& {}
    -\massD\Delta\gamma_i
    -\abs{\massF}^2\Delta\dot{\gamma_i}
    -\left\{\massF^\dagger\Delta\dot{A}_i+\Hc\right\} \ ,
\label{mass:tree}
\end{eqnarray}
where $\dot{A}_i$ can be calculated if 
the anomalous dimension $\gamma_i\equiv{}d\ln{}Z_i^{-1}/d\ln\mu^2$ 
is known as a function of the gauge and Yukawa couplings.

According to the above formulas,
the threshold correction to each soft supersymmetry breaking parameter
is basically proportional to the difference 
of the corresponding anomalous dimensions
between the high-energy theory and the effective theory.
This is interesting because
much is known of the supersymmetric anomalous dimensions;
In the so-called NSVZ scheme, 
we have a cerebrated formula \cite{NSVZ}\tocite{rescaling}
relating the renormalized gauge coupling $\alpha\fun{\mu}$
to the holomorphic counterpart $\alpha_{{\mathrm{H}}}\fun{\mu}$ as
\begin{eqnarray}
\alpha^{-1}+\Nc\ln\alpha
  = \alpha^{-1}_{{\mathrm{H}}}+\sum_{i}T_i\ln{}Z_i^{-1/2} \ ,
\label{holomorphy}
\end{eqnarray}
which implies the well-known form of the gauge beta function
\begin{eqnarray}
\beta
\ \equiv\ \mu\frac{d}{d\mu}\alpha
\ ={}-\frac{\alpha^2}{1-\Nc\alpha}
     \left(b+\sum_{i}T_i\gamma_i\right)
	\ ,
\label{NSVZ}
\end{eqnarray}
where $T_i$ is the Dynkin index of $\phi_i$ representation.
Thus one expects that the formulas as above 
can be used to 
estimate the size of soft parameters
even in the models involving
strongly-coupled gauge dynamics \cite{NS}\tocite{inducedYukawa}.

In the following we will examine our procedure in some concrete models.
Let us first suppose that 
the supersymmetry breaking is dominated by gauge mediation.
This implies that
the soft terms vanish above the messenger scale and thus
\begin{eqnarray}
\left(\massF\right)^{\mathrm{GM}}
 = 	\frac{1}{1-\gamma_M'\fun{\M}}
	\left(\frac{F}{M_{\mathrm{H}}}\right) \ ,
\label{threshold:F:GMSB}
\end{eqnarray}
Substituting this into the formula \eq{gaugino:tree}
and using the beta function \eq{NSVZ}, we find
\begin{eqnarray}
m_g^{{\rm GM}}\fun{\M}
 =\Delta{}m_g\fun{\M}
 ={}-\frac{\alpha\fun{\M}}{1-\Nc\alpha\fun{\M}}
     \left[
	\frac{\Nm}{2}
         +\Nf\,\frac{\Delta\gamma_i\fun{\M}}
                    {1-\gamma_M'\fun{\M}}
     \right]
     \frac{F}{M_{\mathrm{H}}} \ . \qquad
\label{screening}
\end{eqnarray}
When there is no direct superpotential coupling 
between the visible and messenger sectors,
$\Delta\gamma_i$ is already a two-loop quantity,
and thus the gaugino mass at three-loop
is independent from the anomalous dimensions
of the heavy matter fields, $\gamma'_M$.
This is precisely the claim 
of `gaugino screening' theorem \cite{continuation,NLOgaugino}
in the context of GMSB.

Similarly we can check the formulas \eqs{Aterm:tree}{mass:tree}
for the $A$-term and scalar mass;
When we confine ourselves to the GMSB case,
it is bound to be the truth that
these formulas reproduce the correct results.
Actually we have checked our formulas by considering 
the models in which the messenger fields have 
a direct superpotential coupling to the light sector,
as in the model discussed in Ref.~\cite{GiudiceRattazzi}.

As the second example,
we consider the minimal SU(5) model, for which
the one-loop threshold corrections 
at the SU(5)-breaking scale 
were calculated for the gaugino masses \cite{HMG} and 
for the $A$-terms and scalar masses \cite{PolonskyPomarol}.
The difference from the GMSB case is that
the heavy fields have non-zero soft parameters.
In this case,
the model contains as heavy chiral multiplets
the $\mathrm{SU}(3)\times\mathrm{SU}(2)\times\mathrm{U}(1)$ 
components $\Sigma_8$, $\Sigma_3$, $\Sigma_1$  
of the adjoint Higgs $\Sigma\fun{\mathbf{24}}$
as well as the colored Higgses 
$H_{\mathrm{c}}$ and $\ol{H}_{\mathrm{c}}$.
Since our $B_{\mathrm{R}}$ when truncated to one-loop
is just the renormalized $B$-parameter,
the soft parameters of these heavy fields are calculated,
in the notation of Ref.~\cite{PolonskyPomarol}, to be
$B_{\mathrm{R}}=B_\Sigma$ for $\Sigma_{8,3,1}$ and
$B_{\mathrm{R}}=B_H$ for $H_{\mathrm{c}}$ and $\ol{H}_{\mathrm{c}}$.
As a result, it is straightforward to check that
our formula for $\Delta{}m_g$ reproduces
the last terms of Eqs.~(20)--(22) in Ref.~\cite{HMG}.
The other terms are the contributions 
from heavy vector supermultiplets, 
can be reproduced in a similar manner.
The details will be discussed elsewhere.

It should be noticed, however, that
the $A$-parameters (trilinear scalar couplings) are \textit{not} 
contained in \Eq{Aterm:tree}.
To incorporate the $A$-parameters of the heavy fields, 
we improve the matching condition \eq{matching:Z:tree} 
by including the higher order corrections.
At one-loop level, the improved matching conditions take the form
\begin{eqnarray}
\ln Z_i^{-1}(\M)
  &=& \ln Z_i'^{-1}(\M) 
   + \left(\Delta\gamma_i\right)_{\mbox{\scriptsize 1-loop}} \ .
\label{matching:Z:loop}
\end{eqnarray}
When the light field $\phi_i$ has a direct superpotential coupling $y'$
to heavy field(s), 
\begin{eqnarray}
\left(\Delta\gamma_i\right)_{\mbox{\scriptsize 1-loop}}
 &=& -\sum_{y}d_i^{(y)}
	\frac{\norm{y'(\M)}^2}{8\pi^2} \ ,
\label{gamma:oneloop}
\end{eqnarray}
where $d_i^{(y)}$ is a positive constant.
Then extending the above matching conditions into superspace gives
\begin{eqnarray}
\Delta A_i\fun{\M}
  &=& 	\sum_{y}\left[B_{\rm R}-A_y'\right]
	d_i^{(y)}\frac{\norm{y'}^2}{8\pi^2} \ ,
\label{A:improved}\\
\Delta m_i^2\fun{\M}
  &=&   \sum_{y}\left[
	Y_{\rm R} + \left(B_{\rm R}^\dag A_y' + \Hc\right)
	-\left( X_y' + \norm{A_y'}^2 \right)
	\right]
	d_i^{(y)}\frac{\norm{y'}^2}{8\pi^2} \ .
\label{mass:improved}
\end{eqnarray}
In each of these equations,
the last term corresponds to the modification
due to the one-loop matching \eq{matching:Z:loop},
while the terms containing $B_{\mathrm{R}}$ or $Y_{\mathrm{R}}$
come from the previous formula \eqs{Aterm:tree}{mass:tree}
based on the tree-level matching.

Let us come back to the SU(5) example.
At each mass threshold of 
$H_{\mathrm{c}}$, $\Sigma_{3}$ and $\Sigma_{1}$, 
we apply the formula \eq{A:improved} to the $A$-parameter $A_t$ 
corresponding to the top Yukawa coupling $y_t$.
We retain only $y_t$ and also the coupling $\lambda$ 
of $\Sigma\fun{\mathbf{24}}$ to the fundamental Higgses.
The latter coupling $\lambda$ appears 
in the anomalous dimension \eq{gamma:oneloop}
of the up-type weak Higgs $\ol{H}_{\mathrm{w}}$
through the superpotential
\begin{eqnarray}
\W
 &=&
 \sqrt{2}\lambda
	 \left[
	 H_{\mathrm{c}}
	 \Sigma\fun{\mathbf{3}^*,\mathbf{2}}\ol{H}_{\mathrm{w}}
	+H_{\mathrm{w}}\Sigma\fun{\mathbf{1},\mathbf{3}}
	 \ol{H}_{\mathrm{w}}
	+\frac{1}{2}\sqrt{\frac{3}{5}}
	 H_{\mathrm{w}}\Sigma\fun{\mathbf{1},\mathbf{1}}
	 \ol{H}_{\mathrm{w}}
	\right] \ ,
\end{eqnarray}
where 
$\mathrm{SU}(3)\times\mathrm{SU}(2)$ 
quantum numbers have been indicated,
and $\Sigma\fun{\mathbf{1},\mathbf{3}}=T^a\Sigma_{3}^a$
and $\Sigma\fun{\mathbf{1},\mathbf{1}}=\Sigma_{1}$.
Then our formula yields
\begin{eqnarray}
\Delta A_t\fun{M_{H_{\mathrm{c}}}}
  &=&
	 \frac{3}{2}\left(B_H-A_\lambda\right)
	 \frac{\norm{\lambda}^2}{8\pi^2}
	+\frac{3}{2}\left(B_H-A_t\right)
	 \frac{\norm{y_t}^2}{8\pi^2}
	 \ , \qquad
\label{A1}
\\
\Delta A_t\fun{M_{\Sigma_{3}}}
  &=& 	 \frac{3}{4}\left(B_\Sigma-A_\lambda\right)
	 \frac{\norm{\lambda}^2}{8\pi^2} \ ,
\\
\Delta A_t\fun{M_{\Sigma_{1}}}
  &=& 	\frac{3}{20}\left(B_\Sigma-A_\lambda\right)
	\frac{\norm{\lambda}^2}{8\pi^2} \ ,
\label{A3}
\end{eqnarray}
which are consistent\footnote{
The $B$-term contributions, 
which were missing in Ref.~\cite{PolonskyPomarol},
are important in the GMSB case.
}
with Eq.~(B15) in Ref.~\cite{PolonskyPomarol}.
Similarly we find that
the formula \eq{mass:improved} for soft scalar masses
gives consistent results,
confirming our procedure based on
the improved matching condition \eq{matching:Z:loop}.

In this letter,
we have discussed the structure of threshold corrections
to soft supersymmetry breaking parameters
that arise when heavy chiral superfields are integrated out.
A simple application of the superfield coupling formalism
enables us to express the soft threshold effects 
in terms of supersymmetric anomalous dimensions.
We have also argued that
the $A$-parameters of the heavy fields can be incorporated 
by improving the matching condition of matter wavefunctions.
The resultant formulas have been confirmed explicitly at one-loop level,
but in principle, the procedure can be generalized to higher orders.
Such `exact' threshold formulas
have a possible application to models involving
strongly-coupled gauge dynamics \cite{NS}\tocite{inducedYukawa},
and deserve for further study.

{}Finally we add a comment
on the matching condition for the gauge coupling.
If the one-loop matching condition is 
to be used for matter wavefunctions $Z_i$,
the matching condition for the gauge coupling should also be improved.
However the resultant modification will appear 
from two-loop level,
as can be seen from the Shifman-Vainshtein formula \eq{holomorphy}.
Accordingly,
we can determine the two-loop matching relation for the gauge couplings,
from which the two-loop threshold correction
to the gaugino masses can be calculated by our procedure.
It would be interesting to check such two-loop relations explicitly.

\section*{Acknowledgments}
The authors thank Tatsuo~Kobayashi, Haruhiko~Terao
and Yoshihisa~Yamada for discussions.
They also thank the Yukawa Institute for Theoretical Physics at
Kyoto University, where a portion of this work was done during
the YITP-W-06-06. 
H.~N.\ and K.~Y.\ 
are supported in part by the Grants-in-Aid for Scientific Research
(No.~16540238, and No.~16081209 and No.~17740150, respectively)
from the Ministry of Education, Science, Sports and Culture, Japan.

%


\begin{thebibliography}{99}
  
\bibitem{spurion}
Y.~Yamada, \PRD{50,1994,3537};\\
L.V.~Avdeev, D.I.~Kazakov and I.N.~Kondrashuk, \NPB{510,1998,289};\\
I.~Jack, and D.R.T.~Jones, \PLB{415,1997,383}.

\bibitem{HisanoShifman}
J.~Hisano and M.~Shifman, \PRD{56,1997,5475}.

\bibitem{Deltaterm}
I.~Jack, D.R.T.~Jones and A.~Pickering, 
\PLB{426,1998,73}; \andvol{432,1998,114};\\
T.~Kobayashi, J.~Kubo and G.~Zoupanos, \PLB{427,1998,291};\\
D.I.~Kazakov and V.N.~Velizhanin, \PLB{485,2000,393}.

\bibitem{GiudiceRattazzi}
G.~Giudice and R.~Rattazzi, \NPB{511,1998,25}.

\bibitem{continuation}
N.~Arkani-Hamed, G.F.~Giudice, M.A.~Luty, R.~Rattazzi, 
\PRD{58,1998,115005}.

\bibitem{matching}
S.~Weinberg, \PLB{91,1980,51};\\
H.~Hall, \NPB{178,1981,75}.

\bibitem{NSVZ} 
V.~Novikov, M.~Shifman, A.~Vainshtein and V.~Zakharov,   
\PLB{166,1986,329}. 

\bibitem{ShifmanVainshtein} 
M.~Shifman and A.~Vainshtein, \NPB{359,1991,571}.

\bibitem{rescaling}
N.~Arkani-Hamed and H.~Murayama, 
\JHEP{06,2000,030}.

\bibitem{NS}
A.E.~Nelson and M.J.~Strassler, \JHEP{09,2000,030};
ibid. \andvol{07,2002,021}.
\bibitem{KT}
T.~Kobayashi and H.~Terao, \PRD{64,2001,075003};\\
T.~Kobayashi, H.~Nakano, T.~Noguchi and H.~Terao, 
\PRD{66,2002,095011}.

\bibitem{KNT}
T.~Kobayashi, H.~Nakano and H.~Terao, \PRD{65,2002,015006};\\
T.~Kobayashi, H.~Nakano, T.~Noguchi and H.~Terao, 
\JHEP{02,2003,022}.

\bibitem{LS}
M.A.~Luty and R.~Sundrum, 
\PRD{65,2002,066004}; \andvol{67,2003,045007}.


\bibitem{SchmaltzSundrum}
M.~Schmaltz and R.~Sundrum, \JHEP{11,2006,011}.

\bibitem{IINSY}
M.~Ibe, K-I.~Izawa, Y.~Nakayama, Y.~Shinbara, and T.~Yanagida,
\PRD{73,2006,015004}; ibid. 
\andvol{73,2006,035012}.

\bibitem{inducedYukawa}
T.~Kobayashi, H.~Nakano and H.~Terao, \PRD{71,2005,115009};\\
T.~Kobayashi, H.~Terao and A.~Tsuchiya, \PRD{74,2006,015002}.

\bibitem{NLOgaugino} 
M.~Picariello and A.~Strumia, \NPB{529,1998,81}.

\bibitem{HMG}
J.~Hisano, H.~Murayama and T.~Goto,  \PRD{49,1994,1446}.

\bibitem{PolonskyPomarol}
N.~Polonsky and A.~Pomarol, \PRD{51,1995,6532}.

\end{thebibliography}
\end{document}